\documentclass{article}

\usepackage{arxiv}

\usepackage[utf8]{inputenc} 
\usepackage[T1]{fontenc}    
\usepackage{hyperref}       
\usepackage{url}            
\usepackage{booktabs}       
\usepackage{amsfonts}       
\usepackage{nicefrac}       
\usepackage{microtype}      
\usepackage{lipsum}		

\usepackage{graphicx}
\usepackage{amssymb}
\usepackage{mathtools}
\usepackage{multirow}
\usepackage{booktabs}
\usepackage{bm}

\DeclarePairedDelimiter\abs{\lvert}{\rvert}%

\DeclareMathOperator*{\argmin}{arg\,min}

\usepackage{amsthm}
\newtheorem{thm}{Lemma}[section]

\title{Minimum message length inference of the Weibull distribution with complete and censored data}

\author{ \href{https://orcid.org/0000-0003-3017-0871}{\includegraphics[scale=0.06]{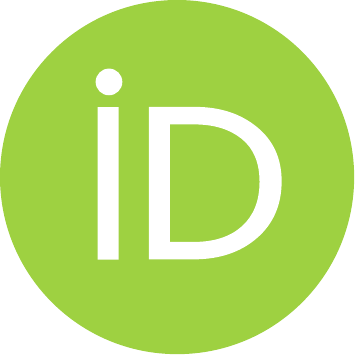}\hspace{1mm}Enes Makalic} \\
	Melbourne School of Population and Global Health\\
	University of Melbourne\\
	Carlton, VIC 3053 \\
	\texttt{emakalic@unimelb.edu.au} \\
	\And
	\href{https://orcid.org/0000-0000-0000-0000}{\includegraphics[scale=0.06]{orcid.pdf}\hspace{1mm}Daniel F.~Schmidt} \\
	Faculty of Information Technology\\
	Monash University\\
	Clayton, VIC 3168 \\
	\texttt{dschmidt@monash.edu} \\
}

\date{}

\renewcommand{\shorttitle}{MML inference of the Weibull distribution}

\hypersetup{
pdftitle={Minimum message length inference of the Weibull distribution with complete and censored data},
pdfauthor={Enes Makalic, Daniel F.~Schmidt},
pdfkeywords={minimum message length, Weibull distribution, parameter estimation, model selection},
}

\begin{document}
\maketitle

\begin{abstract}
The Weibull distribution, with shape parameter $k>0$ and scale parameter $\lambda>0$, is one of the most popular parametric distributions in survival analysis with complete or censored data. Although inference of the parameters of the Weibull distribution is commonly done through maximum likelihood, it is well established that the maximum likelihood estimate of the shape parameter is inadequate due to the associated large bias when the sample size is small or the proportion of censored data is large. This manuscript demonstrates how the Bayesian information-theoretic minimum message length principle coupled with a suitable choice of weakly informative prior distributions, can be used to infer Weibull distribution parameters given complete data or data with type I censoring. Empirical experiments show that the proposed minimum message length estimate of the shape parameter is superior to the maximum likelihood estimate and appears superior to other recently proposed modified maximum likelihood estimates in terms of Kullback-Leibler risk. Lastly, we derive an extension of the proposed method to data with type II censoring.
\end{abstract}


\section{Introduction}
\label{sec:introduction}
The Weibull distribution is one of the most important probability distributions in analysis of lifetime data. The probability density function and cumulative density function of a Weibull random variable $T$ with shape parameter $k>0$ and scale parameter $\lambda>0$ are 
\begin{align}
\label{eqn:y:complete}
	p_T(t|k, \lambda) = \left(\frac{k}{\lambda^k}\right) t^{k-1} \exp\left(-\left(\frac{t}{\lambda}\right)^k\right), \quad F_T(t | k, \lambda) = 1 - \exp\left(-\left(\frac{t}{\lambda}\right)^k\right), 
\end{align}
respectively. With lifetime data, we often do not observe complete data and instead have joint realisations of the random variables $(Y = y, \Delta = \delta)$ where
\begin{eqnarray}
\label{eqn:y:censored}
	Y &=& \min (T, C) \\
\label{eqn:Delta}	
	\Delta_i &=& {\rm I}(T \leq C) = 
	\begin{cases}
    1, & \text{if } T \leq C \; ({\rm observed\; survival})\\
    0, & \text{if } T > C \; ({\rm observed\; censoring})
	\end{cases}
\end{eqnarray}
where the random variables $T$ and $C$ are assumed to be independent and denote the survival time and the censoring time of an item, respectively. In words, we observe the survival time $T=t$ of an item if it is less than the corresponding censoring time $C = c$ (i.e., $t \leq c$) ; otherwise, we only know that the item survived past time $c$ (i.e., $t > c$). The censoring time may be a fixed constant (say, $c$) or a random variable that may depend on other factors. Given $n$ i.i.d. lifetime data points ${\bf y} = (y_1,\ldots,y_n)$ with or without censoring, an important problem is to estimate the unknown parameters $k$ and $\lambda$ and thus learn about the survival distribution of the items.

The most common approach to parameter estimation is the method of maximum likelihood, where the unknown parameters are set to values that maximise the (log-) likelihood of the observed data. Unfortunately, in the case of the Weibull shape parameter the corresponding maximum likelihood estimate is known to have large bias with both complete and censored data (see, for example, \cite{Ross94, MackisackStillman96, Hirose99}) and this is especially evident for small sample sizes and/or under large amounts of censoring. This manuscript introduces the Bayesian minimum message length (MML) approach to inductive inference and demonstrates how MML can be used to estimate Weibull parameters in both the complete and censored data setting. We show that with an appropriate choice of prior distributions the MML estimate of the shape parameter improves on the maximum likelihood estimate, given censored or complete data, and is competitive with alternative proposals that modify the maximum likelihood estimate to reduce bias. Furthermore, we demonstrate how the MML principle can be used to discriminate between the lognormal and Weibull distributions with censored data. Empirical experiments suggest that model selection with MML is an excellent alternative to commonly used information criteria such as the Bayesian information criterion.
\section{Minimum message length}
\label{sec:mml}
The minimum message length (MML) principle~\cite{WallaceBoulton68, WallaceFreeman87, WallaceDowe99a, Wallace05} is a Bayesian information-theoretic framework for inductive inference that provides a new, unified approach to parameter estimation and model selection. Given data, the key step in applying MML is the computation of the length of a message that describes (encodes) the data, with the assumption that the message comprises two components: 
\begin{enumerate}
\item the \emph{assertion}, encoding of the structure of the model, including all model parameters $\bm{\theta} \in \bm{\Theta} \in \mathbb{R}^p$; and
\item the \emph{detail},  encoding the data $D$ using the model $p(D | \bm{\theta})$ from the assertion. 
\end{enumerate}
The length of the assertion measures the complexity of the model, with simpler models having a shorter assertion compared to more complex models. The length of the detail measures how well the model named in the assertion fits the data; more complex models will have shorter detail lengths compared to simpler models. The length of the combined two-part message, $I(D, \bm{\theta})$, is 
\begin{equation}
\label{eqn:mml:codelength}
I(D, \bm{\theta}) = \underbrace{ I(\bm{\theta}) }_{\rm assertion} + \underbrace{I(D | \bm{\theta})}_{\rm detail} 
\end{equation}
i.e., the sum of the length of the assertion, $I(\bm{\theta})$, and the length of detail, $I(D|\bm{\theta})$. Inference in the MML framework proceeds by finding the model
\begin{equation}
\hat{\bm{\theta}}(D) = \argmin_{\bm{\theta} \in \bm{\Theta}} \left\{ I(D, \bm{\theta}) \right\}
\end{equation}
that minimises the length of the two-part message message. Minimising the MML codelength requires balancing complexity of a model (assertion) with the corresponding fit to the data (detail) with the preferred model being the simplest model that fits the data sufficiently well. A key advantage of MML is that the unit of measurement, the codelength (generally measured in $\log_e$ digits, called nits or nats), is universal in the sense that allows inference and comparison of models with different model structures (e.g., linear regression vs. decision tree) and parameters within a single, unified framework.

Precise computation of codelengths is known to be a NP-hard proble in general. As such, there exist many MML approximations to the codelength (\ref{eqn:mml:codelength})~\cite{WallaceBoulton75,Wallace05}, with the MML87 approximation~\cite{WallaceFreeman87,Wallace05} being the most widely applied due to it's relative computational simplicity. Under suitable regularity conditions, the MML87 codelength approximates (\ref{eqn:mml:codelength}) by
\begin{equation}
\label{eqn:mml87:codelength}
I_{87}(D, \bm{\theta}) = \underbrace{-\log \pi(\bm{\theta}) + \frac{1}{2} \log \abs{J_{\bm{\theta}}(\bm{\theta})} + \frac{p}{2} \log \kappa_p}_{\rm assertion} + \underbrace{\frac{p}{2} - \log p(D|\bm{\theta})}_{\rm detail}
\end{equation}
where $\pi_{\bm{\theta}}(\bm{\theta})$ is the prior distribution of the parameters $\bm{\theta}$, $\abs{J_{\bm{\theta}}(\bm{\theta})}$ is the determinant of the expected Fisher information matrix, $p(D|\bm{\theta})$ is the likelihood function of the model and $\kappa_p$ is a quantization constant~\cite{ConwaySloane98,AgrellEriksson98}; for small $p$ we have 
\begin{equation}
\kappa_1 = \frac{1}{12}, \quad \kappa_2 = \frac{5}{36 \sqrt{3}}, \quad \kappa_3 = \frac{19}{192 \times  2^{1/3}},
\end{equation}
while, for moderate to large $p$, $\kappa_p$ is well-approximated by~\cite{Wallace05}:
\begin{equation}
\frac{p}{2} (\log \kappa_p + 1) \approx -\frac{p}{2} \log 2\pi + \frac{1}{2} \log p \pi - \gamma,
\end{equation}
where $\gamma \approx 0.5772$ is the Euler--Mascheroni constant. The MML87 approximation is invariant under smooth one-to-one reparametarizations of the likelihood function and is asymptotically equivalent to the Bayesian information criterion (BIC)~\cite{Schwarz78} as $n \to \infty$ with $p>0$ fixed; that is,
\begin{equation}
\label{eqn:mml87:bic}
	I_{87}(D, \bm{\theta}) = - \log p(D|\bm{\theta}) + \frac{p}{2} + O(1)
\end{equation}
where the $O(1)$ term depends on the prior distribution, the Fisher information and the number of parameters $p$. 
%
%
There exist many successful applications of the MML principle in statistics and machine learning literature, including factor analysis~\cite{WallaceFreeman92}, time series~\cite{Schmidt08a,SchmidtMakalic16a}, linear causal models~\cite{WallaceKorb99} and mixture models~\cite{WallaceDowe00,SchmidtMakalic12}), among others.

\section{Complete data}
\subsection{Maximum likelihood estimates}
\label{sec:complete:mle}
Consider first the setting of complete data with no censoring. The negative log-likelihood of data ${\bf y} = (y_1,\ldots,y_n)$  is
\begin{equation}
\label{eqn:complete:nll}
- \log p_T({\bf y} | k, \lambda) = n \log \left(\frac{\lambda^k}{k}\right) - (k-1) \left(\sum_{i=1}^n \log y_i\right) + \sum_{i=1}^n \left(\frac{y_i}{\lambda}\right)^k
\end{equation}
The maximum likelihood (ML) estimates of $k,\lambda$ are 
\begin{equation}
\hat{\lambda}^{k}({\bf y}) = \frac{1}{n} \sum_{i=1}^n y_i^{k}, 
\end{equation}
where $\hat{k}({\bf y})$ is defined implicitly by
\begin{equation}
\label{eqn:mle:kscore}
 \frac{n}{k} + \sum_{i=1}^n \log y_i - \frac{n \sum_i y_i^k \log y_i}{\sum_i y_i^k} = 0
\end{equation}
and must be solved for numerically. While the ML estimate of $\lambda$ is reasonable, the ML estimate of $k$ is known to exhibit large bias and perform poorly in terms of squared error risk, especially for small sample sizes~\cite{MackisackStillman96}. 

Several attempts have been made to construct a modified ML estimate with improved performance. Ross~\cite{Ross94} derives the simple adjustment formula 
\begin{equation}
\label{eqn:mle:kross}
	\hat{k}_{\rm R}({\bf y}) = \left(\frac{n-2}{n-0.68}\right) \hat{k}_{\rm ML}({\bf y})
\end{equation}
for the ML estimate that reduces the bias to typically better than about $0.05$\%, though this adjustment applies to complete data only. Similarly, Hirose~\cite{Hirose99} derives tables with correction coefficients that can be used to obtain modified ML estimates of both $k$ and $\lambda$ with reduced bias. In a somewhat different approach, Yang and Xie~\cite{YangXie03} apply the modified profile likelihood proposed by Cox and Reid~\cite{CoxReid87,CoxReid92} to derive a penalized maximum likelihood estimate of $k$. Specifically, the Yang and Xie estimate of $\lambda$ is equivalent to the ML estimate while the new estimate of the shape parameter $k$ is obtained by numerically solving 
\begin{equation}
\label{eqn:ml:yangxie}
 \frac{n-2}{k} + \sum_{i=1}^n \log y_i - \frac{n \sum_i y_i^k \log y_i}{\sum_i y_i^k} = 0.
\end{equation}
which is similar to (\ref{eqn:mle:kscore}), the only difference being $(n-2)$ in the numerator of the first term. Yang and Xie empirically show that their estimate of $k$ is less biased than the ML estimate and is more efficient than the simple modification (\ref{eqn:mle:kross}) proposed by Ross. In the next section, we show how to derive an MML estimate of the Weibull distribution parameters and demonstrate that the Yang and Xie modified maximum likelihood estimate is an MML87 estimate for a particular prior distribution. 
\subsection{Minimum message length estimates}
\label{sec:complete:mml}
To derive the MML87 codelength~(\ref{eqn:mml87:codelength}) we require the determinant of the expected Fisher information matrix
\begin{equation}
\label{eqn:fisher:complete}
	| J(k, \lambda) | = \frac{n^2 \pi^2}{6 \lambda ^2}, 
\end{equation}
and prior distributions for both parameters. Assuming that $k$ and $\lambda$ are independent a priori, we opt for the half-Cauchy distributions 

\begin{equation}
\label{eqn:mmlprior:complete}
	\pi(k, \lambda) = \pi(k) \pi(\lambda), \quad \pi(k) = \frac{2}{\pi (1 + k^2)}, \quad \pi(\lambda) = \frac{2}{\pi (1 + \lambda^2)}.
\end{equation}
As $\lambda$ is a scale parameter, a heavy tailed distribution like the half-Cauchy is appropriate and recommended in, for example, \cite{PolsonScott12}. Additionally, the half-Cauchy distribution is suitable for the shape parameter $k$ as $k=1$ denotes a fixed (constant) failure rate and  decreasing ($k < 1$) and increasing ($k > 1$) failure rate are assumed equally likely a priori; that is,
\begin{equation}
	\int_0^1 \pi(k) dk = \int_1^\infty \pi(k) dk = \frac{1}{2} .
\end{equation}
The complete MML87 codelength for the Weibull distribution is
\begin{equation}
\label{eqn:mml87:weibull:codelength}
I_{87}(D, k, \lambda) = -\log \left(\frac{4}{\pi^2 (1+k^2)(1+\lambda^2)} \right) + \frac{1}{2} \log \left( \frac{n^2 \pi^2}{6 \lambda ^2} \right) - \log p_T({\bf y} | k, \lambda) + 1 + \log \kappa_2
\end{equation}
where the negative log-likelihood function $- \log p_T({\bf y} | k, \lambda)$ is given in (\ref{eqn:complete:nll}) and $\kappa_2 = 5/(36 \sqrt{3})$ (see Section~\ref{sec:mml}). Unfortunately, with this selection of prior distributions, the MML87 estimates of $k$ and $\lambda$ must be obtained by numerically minimising (\ref{eqn:mml87:weibull:codelength}).

It is straightforward to see that the modified maximum likelihood estimate of Yang and Xie (\ref{eqn:ml:yangxie}) is the MML87 estimate obtained under the prior distribution
\begin{equation}
\pi(k, \lambda) = \pi(k) \pi(\lambda), \quad \pi(k) \propto  \frac{1}{k^2}, \quad \pi(\lambda) \propto \frac{1}{\lambda}.
\end{equation}
which is improper unless lower and upper bound limits are imposed on both the shape and scale parameters. The implied prior distribution for $\lambda$ is the usual scale invariant distribution often used to model a scale parameter while the prior distribution for the shape parameter $k$ is heavy tailed and Cauchy-like asymptotically. As the aforementioned implied prior distributions are similar to (\ref{eqn:mmlprior:complete}) in their behaviour, it is expected that both the Yang and Xie modified maximum likelihood estimate and the MML87 estimate proposed in this manuscript will yield similar parameter estimates with virtually identical properties.
\section{Censored data}
\label{sec:censored}
%
We now examine inference of the Weibull distribution in the presence of Type I fixed as well as random censoring. Consider first the fixed censoring setup where observations are censored after some  period of time $c > 0$. In particular, we observe the lifetime of an item only if $T_i \leq c$, otherwise we observe the censoring time $c$. The likelihood function of $n$ observed data points $D = \{(y_1, \delta_1), \ldots, (y_n, \delta_n)\}$ is
\begin{equation}
	p(D) = \prod_{i=1}^n p_{T}(y_i)^{\delta_i} (1 - F_{T}(y_i))^{1 -\delta_i}
\end{equation}
where $\delta_i = 1$ if the survival time is observed, and $\delta_i = 0$ if the censoring time is observed (see Section~\ref{sec:introduction}). 

In contrast, under random censoring, both the lifetime  $T_i$ and the censoring time $C_i$ are assumed to be mutually independent random variables. Here, the likelihood function of $n$ observed data points $D = \{(y_1, \delta_1), \ldots, (y_n, \delta_n)\}$ can be written as
\begin{equation*}
	p(D) = \left(\prod_{i=1}^n p_{T}(y_i)^{\delta_i} (1 - F_{T}(y_i))^{1 -\delta_i}\right) \left( \prod_{i=1}^n p_C(y_i)^{1-\delta_i} (1 - F_{C}(y_i))^{\delta_i} \right)
\end{equation*}
where $p_T(t|\theta)$ and $F_T(t|\theta)$ denote the probability density and the cumulative density function of the random variable $T$, respectively. We assume the random censoring setup examined in \cite{DanishAslam12}, where both $T_i$ and $C_i$ are Weibull random variables
\begin{equation}
\label{eqn:exponential}
	T_i \sim {\rm Weibull}(\theta, \beta), \quad C_i \sim {\rm Weibull}(\theta, \alpha), \quad i = 1,\ldots,n,
\end{equation}
where $\alpha,\beta>0$ are the scale parameters and $\theta > 0$ is the common shape parameter. 
The joint probability density function of $Y_i = {\rm min}(T_i, C_i)$ and $\Delta_i = I(T_i < C_i)$ is
\begin{equation}
p_{Y,\Delta}(y, \delta | \alpha, \beta, \theta) = \left( \frac{\theta}{\alpha^\theta} \right) \left( \frac{\alpha}{\beta} \right)^{\delta_i \theta} y^{\theta-1} \exp\left(- \left(\frac{1}{\alpha ^{\theta }}+\frac{1}{\beta ^{\theta }}\right) y^{\theta }\right) .
\end{equation}
Next we derive maximum likelihood estimates for the Weibull distribution under type I and random censoring.
\subsection{Maximum likelihood estimates}
\label{sec:censored:mle}
Consider the type I censoring setup as described in Section~\ref{sec:censored}. The likelihood of $n$ data points $D = \{(y_1, \delta_1), \ldots, (y_n, \delta_n)\}$ is
\begin{align}
p(D)  
%
&= \left(\frac{k}{\lambda^k}\right)^d \exp\left(-\frac{1}{\lambda^k} \sum_{i=1}^n y_i^k \right) \prod_{i=1}^n y_i^{\delta_i (k-1)}
\label{eqn:censored:nll}
\end{align}
The maximum likelihood (ML) estimates of $k,\lambda$ are 
\begin{equation}
\hat{\lambda}^{k}({\bf y}) = \frac{1}{d} \sum_{i=1}^n y_i^{k}, 
\end{equation}
where $d = \sum_{i=1}^n \delta_i$ and $\hat{k}({\bf y})$ is given implicitly by
\begin{equation}
\label{eqn:mle:censored:kscore}
 \frac{d}{k} + \sum_{i=1}^n \delta_i \log y_i - \frac{d \sum_i y_i^k \log y_i}{\sum_i y_i^k} = 0 \, .
\end{equation}
The maximum likelihood estimate of $k$ is known to exhibit large bias in small samples and when the proportion of censoring is high. Sirvanci and Yang~\cite{SirvanciYang84} propose the alternative estimate 
\begin{equation}
	\hat{k}^{-1}(D) = \frac{1}{d g(d / n)} \sum_{i=1}^n \delta_i  (\log c - \log y_i) , 
\end{equation}
where the function $g(\cdot)$ given by
\begin{equation}
	g(p) = \log \log (1 - p)^{-1}  - \frac{1}{p} \int_0^p \log \log (1-t)^{-1} \, dt .
\end{equation}
is a bias correction factor for the bias in estimating $1/k$. Sirvanci and Yang derive finite sample properties of this estimate and show that it has high relative efficiency in estimating $1/k$ over a range of censoring levels (10\% -- 90\% censoring) provided $0 < d < n$. Using the same strategy as in the  complete data case (see Section~\ref{sec:complete:mle}), Yang and Xie~\cite{YangXie03} propose a new modified maximum likelihood estimate of the shape parameter $k$ that is obtained by solving 
\begin{equation}
\label{eqn:ml:yangxie:censored}
 \frac{d-1}{k} + \sum_{i=1}^n \delta_i \log y_i - \frac{d \sum_i y_i^k \log y_i}{\sum_i y_i^k} = 0.
\end{equation}
However, this modified profile score function requires $d>1$ to yield a positive estimate for $k$.

Next, we examine the random censoring setup described in Section~\ref{sec:censored}. The likelihood of the data under the random censoring model is
\begin{equation}
p_D(D | \alpha, \beta, \theta) = \left(\frac{\theta }{\alpha^{\theta }}\right)^n \left(\frac{\alpha }{\beta }\right)^{d \theta } \exp\left(- \left(\frac{1}{\alpha ^{\theta }}+\frac{1}{\beta ^{\theta }}\right) \sum_{i=1}^n y_i^{\theta }\right) \prod_{i=1}^n y_i^{\theta - 1}
\end{equation}
where, as before, $d = \sum_{i=1}^n \delta_i$. From this, the maximum likelihood estimates of $(\alpha,\beta)$ are
\begin{equation}
	\hat{\alpha}_{\rm ML} = \left(\frac{\sum_{i=1}^n y_i^\theta}{n-d}\right)^{1/\theta }, \quad \hat{\beta}_{\rm ML} = \left(\frac{\sum_{i=1}^n y_i^\theta}{d}\right)^{1/\theta }
\end{equation}
while the maximum likelihood estimate of $\theta$ must be obtained by numerical optimisation.
%
Clearly, the maximum likelihood estimates $(\hat{\alpha}_{\rm ML}, \hat{\beta}_{\rm ML})$ exist only if $d \in (0, n)$. Alternatively, maximum likelihood estimates may be obtained by noting the following.

\begin{thm}
\label{thm:jointpdf}
The joint probability density function of $(Y_i, \Delta_i)$ can be written as
\begin{equation}
p_{Y,\Delta}(y, \delta | \alpha, \beta, \theta) = p_{\Delta}(\delta | \phi) \, p_{Y}(y | k, \lambda),
\end{equation}
where $\Delta \sim {\rm binom}(n, \phi)$ and $Y \sim {\rm Weibull} (k, \lambda)$ and 
\begin{equation}
\phi = P(T \leq C) = \frac{\alpha^\theta}{\alpha^\theta+\beta^\theta}, \quad k = \theta, \quad \lambda = \frac{\beta}{(1 + (\beta/\alpha)^\theta)^{1/\theta}}.
\end{equation}
\end{thm}

The proof is straightforward and is omitted. By Lemma~\ref{thm:jointpdf} and invariance of the maximum likelihood estimate, the maximum likelihood estimates of $\alpha,\beta$ and $\theta$ can also be obtained from the usual maximum likelihood estimates for the binomial and Weibull distributions 
\begin{equation}
	\hat{\phi}_{\rm ML} = \frac{1}{n} \sum_{i=1}^n \delta_i, \quad \hat{\lambda}^{\hat{k}}_{\rm ML} = \frac{1}{n} \sum_{i=1}^n y_i^{\hat{k}}, 
\end{equation}
where $\hat{k}$ is given implicitly by
\begin{equation}
\frac{1}{n} \sum_{i=1}^n \log y_i + \frac{1}{k} - \frac{\sum_i y_i^k \log y_i}{\sum_i y_i^k} = 0
\end{equation}
and by noting that
\begin{equation}
\label{eqn:param:transform}
	\theta = k, \quad \alpha = \lambda (1 - \phi)^{-1/\theta}, \quad \beta = \lambda \phi^{-1/\theta}.
\end{equation}
These estimates exist only if $\phi_{\rm ML} \in (0, 1)$ or, equivalently, $d \in (0, n)$. 
%
%
%
%
\subsection{Minimum message length estimates}
\label{sec:censored:mml}
We consider first MML inference under the type I censoring setup described in Section~\ref{sec:censored}. Let 
\begin{equation}
    z_c = \left(\frac{c}{\lambda}\right)^k, \quad p = 1 - \exp(-z_c).
\end{equation}
As with the complete data setting, we assume independent half-Cauchy prior distributions (see (\ref{eqn:mmlprior:complete}) for both the shape and the scale parameters. The expected Fisher information matrix with type I censoring is~\cite{WatkinsJohn04}
\begin{equation*}
J(k,\lambda) = n 
\left(
\begin{array}{cc}
\frac{p+2 \gamma^{(1)}(1,z_c) + \gamma^{(2)}(1,z_c)}{k^2} & -\frac{p+\gamma^{(1)}(1,z_c)}{\lambda } \\
-\frac{p+\gamma^{(1)}(1,z_c)}{\lambda } & p \left(\frac{k}{\lambda}\right)^2 
\end{array}
\right)
\end{equation*}
where $\gamma(\cdot,\cdot)$ is the incomplete gamma function
\begin{align*}
    \gamma(z,x) = \int_0^x t^{z-1} \exp(-t) dt, \quad
    \gamma^{(j)}(z,x) = \frac{d^j \gamma(z,x)}{d z^j} .
\end{align*}

%
%
%
%
%
%
The determinant of the expected Fisher information matrix 
\begin{equation}
   | J(k,\lambda) | = \left(\frac{n}{\lambda}\right)^2 (\gamma^{(2)}(1,z_c) p - \gamma^{(1)}(1,z_c)^2) ,
\end{equation}
is clearly a complicated function of the probability of no censoring, $p$. 
%
%
%
%
%
The MML87 codelength for the Weibull distribution with type I censoring is
\begin{equation}
\label{eqn:mml87:censored:codelength}
I_{87}(D, \bm{\theta}) = -\log \left(\frac{4}{\pi^2 (1+k^2)(1+\lambda^2)} \right) + \frac{1}{2} \log | J(k,\lambda) | - \log p_T({\bf y} | k, \lambda) + 1 + \log \kappa_2
\end{equation}
where the negative log-likelihood function $- \log p_T({\bf y} | k, \lambda)$ is given in (\ref{eqn:censored:nll}) and $\kappa_2 = 5/(36 \sqrt{3})$ (see Section~\ref{sec:mml}). As with the complete data case the MML87 estimates of $k$ and $\lambda$ must be obtained by numerically minimising (\ref{eqn:mml87:censored:codelength}).

Consider next the random censoring setup described in Section~\ref{sec:censored} where the lifetime  $T_i$ and the censoring time $C_i$ are mutually independent Weibull random variables with a common shape parameter. From Lemma~\ref{thm:jointpdf}, the joint density of $(Y_i, \Delta_i)$ can be written as a product of a binomial distribution $\Delta \sim (n, \phi)$ and Weibull distribution $Y | \Delta \sim \text{Weibull}(k, \lambda)$. This implies that an MML code for the data $D$ could comprise two messages with the first message encoding the binary censoring indicators $\bm{\delta} = (\delta_1,\ldots,\delta_n)$, followed by another message that encodes the lifetimes ${\bf y} = (y_1,\ldots,y_n)$ given the censoring data $\bm{\delta}$. With this encoding, the total MML codelength for the data $D = \{(y_1, \delta_1), \ldots, (y_n, \delta_n)\}$ is 
\begin{equation}
\label{eqn:cond:codelength}
	I_{87}(D, \alpha, \beta, \theta) = I_{87}(\bm{\delta}, \phi) + I_{87}({\bf y}, k, \lambda |\bm{\delta}),
\end{equation}
where $\phi$ is the probability of observing an uncensored datum. As with maximum likelihood, MML87 is invariant under one-to-one parameter transformations implying that MML87 estimates of $(\alpha, \beta, \theta)$ can be obtained from MML87 estimates of $(\phi, k, \lambda)$ using the relations (\ref{eqn:param:transform}).

The MML87 codelength of the binomial distribution was derived in, for example,  \cite{Wallace05,WallaceDowe00} and, for a uniform prior distribution on $\phi$, is given by
\begin{equation}
\label{eqn:Idelta}
I_{87}(\bm{\delta}, \phi) = -\left(k + \frac{1}{2}\right) \log \phi - \left(n + \frac{1}{2} - k\right) \log (1-\phi) + \frac{1}{2}(1 + \log (n/12) )
\end{equation}
where, as before, $k = (\sum_i \delta_i)$. The minimum of the codelength is at the MML87 estimate 
\begin{equation}
	\hat{\phi}_{87}(\bm{\delta}) = \frac{k + 1/2}{n + 1} .
\end{equation}
The conditional codelength of the surivival times ${\bf y}$ given the censoring indicators $\bm{\delta}$, $I_{87}({\bf y}, k, \lambda | \bm{\delta})$ is simply the MML87 codelength for the Weibull distribution discussed in Section~\ref{sec:complete:mml}. Note that it is of course possible to derive the MML87 joint codelength and construct a single message for the data $D$, similar to the complete data case discussed in Section~\ref{sec:complete:mml}. Due to the invariance of the MML87 codelength, both approaches will yield exactly the same inferences.
\section{Experiments}
\label{sec:experiments}
%
%
%
Numerical experiments were performed to measure the performance of the newly proposed MML87 estimates  compared to the maximum likelihood estimate and the modified maximum likelihood estimate of Yang and Xie~\cite{YangXie03} with  complete (see Section~\ref{sec:experiments:complete}) and type I censored data (see Section~\ref{sec:experiments:censored}). 
\begin{table*}[tb]
\scriptsize
\begin{center}
\begin{tabular}{ccccccccc} 
\toprule
$n$ & $k$ & \multicolumn{3}{c}{Bias} & & \multicolumn{3}{c}{Mean Squared Error} \\
  &   & MLE & MMLE & MML87 & ~ & MLE & MMLE & MML87 \\
\cmidrule{1-9}
\multirow{4}{*}{10}  &  0.5 &  0.085 & {\bf  0.008} &  0.063 &  &  0.038 & {\bf  0.023} &  0.029\\ 
 &  1.0 &  0.168 & {\bf  0.015} &  0.085 &  &  0.152 & {\bf  0.094} &  0.099\\ 
 &  5.0 &  0.850 & {\bf  0.085} &  0.117 &  &  3.836 &  2.352 & {\bf  2.336}\\ 
 & 10.0 &  1.692 & {\bf  0.164} &  0.181 &  & 14.973 &  9.143 & {\bf  9.124}\\ 
\vspace{-2mm} \\ 
\cmidrule{2-9}
\vspace{-2mm} \\ 
\multirow{4}{*}{20}  &  0.5 &  0.038 & {\bf  0.004} &  0.030 &  &  0.012 & {\bf  0.009} &  0.011\\ 
 &  1.0 &  0.076 & {\bf  0.008} &  0.040 &  &  0.048 & {\bf  0.037} &  0.038\\ 
 &  5.0 &  0.371 & {\bf  0.031} &  0.045 &  &  1.194 &  0.927 & {\bf  0.923}\\ 
 & 10.0 &  0.774 & {\bf  0.093} &  0.100 &  &  4.881 &  3.761 & {\bf  3.757}\\ 
\vspace{-2mm} \\ 
\cmidrule{2-9}
\vspace{-2mm} \\ 
\multirow{4}{*}{50}  &  0.5 &  0.015 & {\bf  0.002} &  0.012 &  &  0.004 & {\bf  0.003} &  0.003\\ 
 &  1.0 &  0.029 & {\bf  0.004} &  0.016 &  &  0.015 & {\bf  0.013} &  0.014\\ 
 &  5.0 &  0.143 & {\bf  0.016} &  0.021 &  &  0.366 &  0.329 & {\bf  0.328}\\ 
 & 10.0 &  0.279 & {\bf  0.025} &  0.028 &  &  1.456 &  1.311 & {\bf  1.310}\\ 
\vspace{-3mm} \\ 
\bottomrule
\vspace{+1mm}
\end{tabular}
\caption{Bias and mean squared error for maximum likelihood (MLE), modified maximum likelihood (MMLE) and MML87 estimates of $k$ computed over $10^5$ simulations runs with $\lambda = 1$.\label{tab:results:complete}}
\end{center}
\end{table*}
\subsection{Complete data}
\label{sec:experiments:complete}
The MML87 estimate of the shape parameter $k$ derived in Section~\ref{sec:complete:mml} is now compared to the maximum likelihood (MLE) estimate (\ref{eqn:mle:kscore}) and the modified maximum likelihood (MMLE) estimate (\ref{eqn:ml:yangxie}) using simulated data. In each simulation run, $n$ data points were generated from the model Weibull$(k, \lambda = 1)$ where $n = \{10, 20, 50\}$ and the shape parameter was set to $k \in \{0.5, 1, 5, 10\}$. Given the data, MLE, MMLE and MML87 estimates were computed and compared in terms of bias and mean squared error. For each value of $(k, n)$ $10^5$ simulations were performed and the average bias and mean squared error results are shown in Table~\ref{tab:results:complete} for each estimate.

It is clear that the MMLE and MML87 estimates improve significantly on the maximum likelihood estimate in terms of both bias and mean squared error for each tested value of $(n,k)$. We further note that the MMLE estimate of $k$ is slightly less biased than the proposed MML87 estimate, though the two estimates are virtually indistinguishable in terms of the average mean squared error. As discussed in Section~\ref{sec:complete:mml}, the MMLE estimate is a special case of the MML87 estimator for a particular choice of the prior distribution with complete data, and it is therefore expected that the two estimates will have similar behaviour.
\subsection{Censored data}
\label{sec:experiments:censored}
We also compared the MML87 estimate (see Section~\ref{sec:censored:mml}) to the maximum likelihood estimate (MLE) (\ref{eqn:mle:censored:kscore}) and the modified maximum likelihood estimate (MMLE)  (\ref{eqn:ml:yangxie:censored}) under type I censored data. The experimental setup was identical to that for complete data with the following changes: (i) the proportion of uncensored observations was set to $p \in \{0.3, 0.5, 0.7, 0.9\}$, and (ii) $n \in \{20,30,40\}$ data points were generated during each simulation run. We restricted the experiments to exclude data sets where the number of uncensored observations $d (=\sum_i \delta_i) < 2$, as the MLE and MMLE estimates are not defined for small $d$. In addition to the bias and the mean squared error in estimating the shape parameter, we computed the Kullback--Leibler (KL) divergence~\cite{KullbackLeibler51} between the data generating model and each estimated model (see Appendix~A). The results averaged over $10^5$ simulations runs for each combination of $(n,p,k)$ are shown in Table~\ref{tab:results:censored}.

We again observe that the MLE estimate of $k$ is strongly biased particularly for small $k$ and $p$. While the MMLE is less biased than the proposed MML87 estimate, the MML87 estimate achieves smaller mean squared error and smaller KL divergence compared to the MMLE in all experiments. Additionally, we observe that the KL divergence for the MMLE model is similar to the MLE model, despite the significant reduction in bias of estimating the shape parameter $k$ achieved by the MMLE. Clearly the proposed MML87 estimate is an improvement over the MLE and highly competitive against estimators that are primarily designed to reduce bias in the MLE, such as the one proposed by Yang and Xie~\cite{YangXie03}.

\begin{table*}[tbph]
\scriptsize
\begin{center}
\begin{tabular}{cccccccccccccccccc} 
\toprule
$n$ & $p$ & $k$ & \multicolumn{3}{c}{Bias} & & \multicolumn{3}{c}{Mean Squared Error} & & \multicolumn{3}{c}{KL Divergence} \\
    &     &     & MLE & MMLE & MML87 & ~ & MLE & MMLE & MML87 & ~ & MLE & MMLE & MML87 \\
\cmidrule{1-14}
\multirow{16}{*}{20} & \multirow{4}{*}{ 0.3} &  0.5 &  0.114 & {\bf  0.002} &  0.070 &  &  0.158 &  0.077 & {\bf  0.040} &  &  0.069 &  0.060 & {\bf  0.042}\\ 
 & &  1.0 &  0.055 & {\bf  0.005} &  0.038 &  &  0.042 &  0.031 & {\bf  0.026} &  &  0.060 &  0.056 & {\bf  0.043}\\ 
 & &  5.0 &  0.037 & {\bf  0.006} &  0.021 &  &  0.020 &  0.017 & {\bf  0.016} &  &  0.057 &  0.054 & {\bf  0.044}\\ 
 & & 10.0 &  0.019 & {\bf -0.001} &  0.006 &  &  0.011 &  0.010 & {\bf  0.010} &  &  0.048 &  0.046 & {\bf  0.039}\\ 
 & \multirow{4}{*}{ 0.5} &  0.5 &  0.114 & {\bf  0.002} &  0.070 &  &  0.158 &  0.077 & {\bf  0.040} &  &  0.069 &  0.060 & {\bf  0.042}\\ 
 & &  1.0 &  0.055 & {\bf  0.005} &  0.038 &  &  0.042 &  0.031 & {\bf  0.026} &  &  0.060 &  0.056 & {\bf  0.043}\\ 
 & &  5.0 &  0.037 & {\bf  0.006} &  0.021 &  &  0.020 &  0.017 & {\bf  0.016} &  &  0.057 &  0.054 & {\bf  0.044}\\ 
 & & 10.0 &  0.019 & {\bf -0.001} &  0.006 &  &  0.011 &  0.010 & {\bf  0.010} &  &  0.048 &  0.046 & {\bf  0.039}\\ 
 & \multirow{4}{*}{ 0.7} &  0.5 &  0.114 & {\bf  0.002} &  0.070 &  &  0.158 &  0.077 & {\bf  0.040} &  &  0.069 &  0.060 & {\bf  0.042}\\ 
 & &  1.0 &  0.055 & {\bf  0.005} &  0.038 &  &  0.042 &  0.031 & {\bf  0.026} &  &  0.060 &  0.056 & {\bf  0.043}\\ 
 & &  5.0 &  0.037 & {\bf  0.006} &  0.021 &  &  0.020 &  0.017 & {\bf  0.016} &  &  0.057 &  0.054 & {\bf  0.044}\\ 
 & & 10.0 &  0.019 & {\bf -0.001} &  0.006 &  &  0.011 &  0.010 & {\bf  0.010} &  &  0.048 &  0.046 & {\bf  0.039}\\ 
 & \multirow{4}{*}{ 0.9} &  0.5 &  0.114 & {\bf  0.002} &  0.070 &  &  0.158 &  0.077 & {\bf  0.040} &  &  0.069 &  0.060 & {\bf  0.042}\\ 
 & &  1.0 &  0.055 & {\bf  0.005} &  0.038 &  &  0.042 &  0.031 & {\bf  0.026} &  &  0.060 &  0.056 & {\bf  0.043}\\ 
 & &  5.0 &  0.037 & {\bf  0.006} &  0.021 &  &  0.020 &  0.017 & {\bf  0.016} &  &  0.057 &  0.054 & {\bf  0.044}\\ 
 & & 10.0 &  0.019 & {\bf -0.001} &  0.006 &  &  0.011 &  0.010 & {\bf  0.010} &  &  0.048 &  0.046 & {\bf  0.039}\\ 
\vspace{-2mm} \\ 
\cmidrule{2-14}
\vspace{-2mm} \\ 
\multirow{16}{*}{30} & \multirow{4}{*}{ 0.3} &  0.5 &  0.067 & {\bf  0.002} &  0.048 &  &  0.059 &  0.039 & {\bf  0.026} &  &  0.042 &  0.039 & {\bf  0.028}\\ 
 & &  1.0 &  0.035 & {\bf  0.003} &  0.024 &  &  0.020 &  0.017 & {\bf  0.016} &  &  0.038 &  0.036 & {\bf  0.029}\\ 
 & &  5.0 &  0.023 & {\bf  0.004} &  0.013 &  &  0.012 &  0.010 & {\bf  0.010} &  &  0.036 &  0.035 & {\bf  0.030}\\ 
 & & 10.0 &  0.016 & {\bf  0.003} &  0.008 &  &  0.007 &  0.007 & {\bf  0.007} &  &  0.034 &  0.033 & {\bf  0.029}\\ 
 & \multirow{4}{*}{ 0.5} &  0.5 &  0.067 & {\bf  0.002} &  0.048 &  &  0.059 &  0.039 & {\bf  0.026} &  &  0.042 &  0.039 & {\bf  0.028}\\ 
 & &  1.0 &  0.035 & {\bf  0.003} &  0.024 &  &  0.020 &  0.017 & {\bf  0.016} &  &  0.038 &  0.036 & {\bf  0.029}\\ 
 & &  5.0 &  0.023 & {\bf  0.004} &  0.013 &  &  0.012 &  0.010 & {\bf  0.010} &  &  0.036 &  0.035 & {\bf  0.030}\\ 
 & & 10.0 &  0.016 & {\bf  0.003} &  0.008 &  &  0.007 &  0.007 & {\bf  0.007} &  &  0.034 &  0.033 & {\bf  0.029}\\ 
 & \multirow{4}{*}{ 0.7} &  0.5 &  0.067 & {\bf  0.002} &  0.048 &  &  0.059 &  0.039 & {\bf  0.026} &  &  0.042 &  0.039 & {\bf  0.028}\\ 
 & &  1.0 &  0.035 & {\bf  0.003} &  0.024 &  &  0.020 &  0.017 & {\bf  0.016} &  &  0.038 &  0.036 & {\bf  0.029}\\ 
 & &  5.0 &  0.023 & {\bf  0.004} &  0.013 &  &  0.012 &  0.010 & {\bf  0.010} &  &  0.036 &  0.035 & {\bf  0.030}\\ 
 & & 10.0 &  0.016 & {\bf  0.003} &  0.008 &  &  0.007 &  0.007 & {\bf  0.007} &  &  0.034 &  0.033 & {\bf  0.029}\\ 
 & \multirow{4}{*}{ 0.9} &  0.5 &  0.067 & {\bf  0.002} &  0.048 &  &  0.059 &  0.039 & {\bf  0.026} &  &  0.042 &  0.039 & {\bf  0.028}\\ 
 & &  1.0 &  0.035 & {\bf  0.003} &  0.024 &  &  0.020 &  0.017 & {\bf  0.016} &  &  0.038 &  0.036 & {\bf  0.029}\\ 
 & &  5.0 &  0.023 & {\bf  0.004} &  0.013 &  &  0.012 &  0.010 & {\bf  0.010} &  &  0.036 &  0.035 & {\bf  0.030}\\ 
 & & 10.0 &  0.016 & {\bf  0.003} &  0.008 &  &  0.007 &  0.007 & {\bf  0.007} &  &  0.034 &  0.033 & {\bf  0.029}\\ 
\vspace{-2mm} \\ 
\cmidrule{2-14}
\vspace{-2mm} \\ 
\multirow{16}{*}{40} & \multirow{4}{*}{ 0.3} &  0.5 &  0.047 & {\bf  0.002} &  0.035 &  &  0.033 &  0.025 & {\bf  0.018} &  &  0.029 &  0.028 & {\bf  0.021}\\ 
 & &  1.0 &  0.025 & {\bf  0.002} &  0.017 &  &  0.014 &  0.012 & {\bf  0.011} &  &  0.027 &  0.026 & {\bf  0.022}\\ 
 & &  5.0 &  0.017 & {\bf  0.003} &  0.009 &  &  0.008 &  0.008 & {\bf  0.007} &  &  0.027 &  0.026 & {\bf  0.023}\\ 
 & & 10.0 &  0.014 & {\bf  0.004} &  0.008 &  &  0.005 &  0.005 & {\bf  0.005} &  &  0.026 &  0.026 & {\bf  0.023}\\ 
 & \multirow{4}{*}{ 0.5} &  0.5 &  0.047 & {\bf  0.002} &  0.035 &  &  0.033 &  0.025 & {\bf  0.018} &  &  0.029 &  0.028 & {\bf  0.021}\\ 
 & &  1.0 &  0.025 & {\bf  0.002} &  0.017 &  &  0.014 &  0.012 & {\bf  0.011} &  &  0.027 &  0.026 & {\bf  0.022}\\ 
 & &  5.0 &  0.017 & {\bf  0.003} &  0.009 &  &  0.008 &  0.008 & {\bf  0.007} &  &  0.027 &  0.026 & {\bf  0.023}\\ 
 & & 10.0 &  0.014 & {\bf  0.004} &  0.008 &  &  0.005 &  0.005 & {\bf  0.005} &  &  0.026 &  0.026 & {\bf  0.023}\\ 
 & \multirow{4}{*}{ 0.7} &  0.5 &  0.047 & {\bf  0.002} &  0.035 &  &  0.033 &  0.025 & {\bf  0.018} &  &  0.029 &  0.028 & {\bf  0.021}\\ 
 & &  1.0 &  0.025 & {\bf  0.002} &  0.017 &  &  0.014 &  0.012 & {\bf  0.011} &  &  0.027 &  0.026 & {\bf  0.022}\\ 
 & &  5.0 &  0.017 & {\bf  0.003} &  0.009 &  &  0.008 &  0.008 & {\bf  0.007} &  &  0.027 &  0.026 & {\bf  0.023}\\ 
 & & 10.0 &  0.014 & {\bf  0.004} &  0.008 &  &  0.005 &  0.005 & {\bf  0.005} &  &  0.026 &  0.026 & {\bf  0.023}\\ 
 & \multirow{4}{*}{ 0.9} &  0.5 &  0.047 & {\bf  0.002} &  0.035 &  &  0.033 &  0.025 & {\bf  0.018} &  &  0.029 &  0.028 & {\bf  0.021}\\ 
 & &  1.0 &  0.025 & {\bf  0.002} &  0.017 &  &  0.014 &  0.012 & {\bf  0.011} &  &  0.027 &  0.026 & {\bf  0.022}\\ 
 & &  5.0 &  0.017 & {\bf  0.003} &  0.009 &  &  0.008 &  0.008 & {\bf  0.007} &  &  0.027 &  0.026 & {\bf  0.023}\\ 
 & & 10.0 &  0.014 & {\bf  0.004} &  0.008 &  &  0.005 &  0.005 & {\bf  0.005} &  &  0.026 &  0.026 & {\bf  0.023}\\ 
\vspace{-3mm} \\ 
\bottomrule
\vspace{+1mm}
\end{tabular}
\caption{Bias, mean squared error and Kullback--Leibler (KL) divergence for maximum likelihood (MLE), modified maximum likelihood (MMLE) and MML87 estimates of $k$ computed over $10^5$ simulations runs with $\lambda = 1$; $p$ denotes the proportion of uncensored observations. \label{tab:results:censored}}
\end{center}
\end{table*}

\section{Discussion}
The minimum message length (MML) principle unifies parameter estimation and model selection within the same framework. This manuscript demonstrates how the MML framework of inductive inference can be applied to the Weibull distribution with complete and censored data. By minimising a single inferential quantity, the codelength, we obtain new parameter estimates that have advantages (in terms of bias and mean squared estimation error) over the usual maximum likelihood estimates. Further, the same codelength can be used to discriminate between competing models such as the Weibull distribution or the lognormal distribution. Using MML for model selection has advantages over alternative popular model selection criteria, such as the Bayesian information criterion, which we demonstrate in the next section. 

\subsection{Model selection}
The task is to infer whether observed data was generated by a Weibull distribution or by a lognormal distribution~\cite{SiswadiQuesenberry82,UpadhyayPeshwani03,KimYum08}. To use MML to discriminate between competing models, we simply need the codelength of the data under each model. For complete data with no censoring, the probability density function of the lognormal distribution with mean $\mu \in \mathbb{R}$ and standard deviation $\sigma > 0$ is
\begin{equation}
    p(y | \mu, \sigma) = \frac{1}{\sqrt{2 \pi } \sigma  y} \exp\left(-\frac{(\log (y)-\mu )^2}{2 \sigma ^2}\right) .
\end{equation}
The negative log-likelihood for data ${\bf y}$ is 
\begin{equation}
\label{eqn:nll:logn:complete}
   -\log p_T({\bf y} | \mu, \sigma) = \frac{n}{2} \log (2\pi) + n \log \sigma + \sum_{i=1}^n \log y_i + \frac{1}{2\sigma^2} \sum_{i=1}^n (\log y_i - \mu)^2 .
\end{equation}
The determinant of the expected Fisher information for the lognormal model is well-known
\begin{equation}
\label{eqn:fisher:logn:complete}
    |J(\mu,\sigma)| = \frac{2 n^2}{\sigma ^4} .
\end{equation}
Similar to Section~\ref{sec:complete:mml}, we select heavy-tailed prior distributions for both parameters 
\begin{equation}
\label{eqn:priors:logn}
    \pi(\mu,\sigma) = \pi(\mu) \pi(\sigma), \quad \pi(\mu) = \frac{1}{\pi (1 + \mu^2)}, \quad \pi(\sigma) = \frac{2}{\pi (1 + \sigma^2)}.
\end{equation}
Substituting (\ref{eqn:nll:logn:complete}), (\ref{eqn:fisher:logn:complete}) and (\ref{eqn:priors:logn}) into (\ref{eqn:mml87:codelength}) yields the MML87 codelength for the lognormal distribution. Due to this choice of prior distributions, the MML87 estimates of $\mu$ and $\sigma$ must be obtained numerically. To determine whether observed data follows the Weibull or the lognormal distribution, we compute the codelength of the data under each model and select the model with the smallest codelength. 

An experiment was setup to compare the MML87 model selection performance against the commonly used Bayesian information criterion (BIC)~\cite{Schwarz78} and the scale transformation maximal invariant statistic (SI)~\cite{QuesenberryKent82}. A comparison of BIC and SI on discriminating between the Weibull and lognormal models was examined   in~\cite{KimYum08}. Similar to the experimental setup in~\cite{KimYum08}, we generated $n \in (10,25,50,100,200)$ data points from either the Weibull(1,1) or the Lognormal(1,1) model, as both BIC and SI are invariant under scale and shape transformations. Each method was then asked to select the best fitting model for the observed data and the experiment was repeated for $10^5$ iterations. The performance of each method was measured in terms of probability of correct selection and the results are shown in Table~\ref{tab:results:mdl:complete}. 

While all three methods tested performed similarly for medium to large sample sizes, the average accuracy of MML87 is significantly higher compared to BIC and SI under small sample sizes. Additionally, while the SI statistic tended to favour the lognormal distribution, no such preference was observed for MML or BIC.

\begin{table*}[tb]
\scriptsize
\begin{center}
\begin{tabular}{cccccccccccc} 
\toprule
$n$ & \multicolumn{3}{c}{${\bf y} \sim$ Weibull} & & \multicolumn{3}{c}{${\bf y} \sim$ Lognormal} & & \multicolumn{3}{c}{Average accuracy}\\
  &   MML87 & BIC & SI & ~ & MML87 & BIC & SI & ~ & MML87 & BIC & SI \\
\cmidrule{1-12}
10  & {\bf  0.738} &  0.677 &  0.596 & &  0.714 &  0.663 & {\bf  0.742} & & {\bf  0.726} &  0.670 &  0.669\\ 
25  & {\bf  0.838} &  0.807 &  0.783 & &  0.826 &  0.803 & {\bf  0.828} & & {\bf  0.832} &  0.805 &  0.806\\ 
50  & {\bf  0.917} &  0.904 &  0.894 & & {\bf  0.913} &  0.904 &  0.913 & & {\bf  0.915} &  0.904 &  0.904\\ 
100  & {\bf  0.975} &  0.972 &  0.970 & & {\bf  0.973} &  0.971 &  0.973 & & {\bf  0.974} &  0.971 &  0.971\\ 
200  & {\bf  0.997} &  0.997 &  0.997 & & {\bf  0.998} &  0.997 &  0.998 & & {\bf  0.997} &  0.997 &  0.997\\ 
\vspace{-3mm} \\ 
\bottomrule
\vspace{+1mm}
\end{tabular}
\caption{Probability of correctly selecting the data generating model for MML87, BIC and SI computed over $10^5$ simulation runs with complete data only. \label{tab:results:mdl:complete}}
\end{center}
\end{table*}

We can of course use the MML principle to discriminate between the Weibull and lognormal distributions based on type I censored data. In this case, the negative log-likelihood function of the data is
\begin{eqnarray}
    -\log p(D|\mu,\sigma) &=& d \log \sigma + \frac{d}{2} \log (2\pi) + \sum_{i=1}^n \delta_i \log(y_i) + \frac{1}{2\sigma^2} \sum_{i=1}^n \delta_i (\log(y_i) - \mu)^2 \nonumber \\
    &&- (n-d)\log\left(1 - \Phi\left(\frac{\log(c) - \mu}{\sigma}\right)\right)~\label{eqn:nll:logn:typeI}
\end{eqnarray}
where $\Phi(\cdot)$ is cumulative density function of the standard normal distribution. Let
\begin{eqnarray}
z = \left(\frac{\log(c) - \mu}{\sigma}\right), \quad p = \Phi(z), \quad M = \Phi^{-1}(p),
\end{eqnarray}
where $z$ is the standardised censoring point and $\Phi^{-1}(\cdot)$ is the inverse cumulative density function of the standard normal distribution. The expected Fisher information matrix is
\begin{equation*}
    J(\mu,\sigma) = \frac{n}{\sigma^2} 
\left(
\begin{array}{cc}
 \frac{e^{-M^2}}{2 \pi (1 - p)}-\frac{e^{-\frac{M^2}{2}} M}{\sqrt{2 \pi }}+p & \frac{e^{-M^2} M}{2 \pi(1 - p)}-\frac{e^{-\frac{M^2}{2}} \left(M^2+1\right)}{\sqrt{2 \pi }} \\
 \frac{e^{-M^2} M}{2 \pi (1 - p)}-\frac{e^{-\frac{M^2}{2}} \left(M^2+1\right)}{\sqrt{2 \pi }} & \frac{e^{-M^2} M^2}{2 \pi(1 -p)}-\frac{e^{-\frac{M^2}{2}} \left(M^3+M\right)}{\sqrt{2 \pi }}+2 p \\
\end{array}
\right)
\end{equation*}
with determinant $|J(\mu,\sigma)|$ given by
\begin{equation}
   \frac{2 n^2}{\sigma^4} \left[\frac{e^{-M^2} \left(M^2 (1-2 p)-3 p+1\right)}{4 \pi  (p-1)}+\frac{e^{-\frac{1}{2} \left(3 M^2\right)} M}{4\sqrt{2} \pi ^{3/2} (1- p)}-\frac{e^{-\frac{M^2}{2}} M \left(M^2+3\right) p}{2\sqrt{2 \pi }}+ p^2\right].
\label{eqn:fisher:logn:typeI}  
\end{equation}
which is equal to the determinant of the expected Fisher information matrix for complete data multiplied by a correction factor that takes into account the proportion of censoring. We use the same prior distributions for the parameters as in the case of complete data. 

The MML87 codelength for the lognormal distribution with type I censoring is obtained by substituting (\ref{eqn:nll:logn:typeI}), (\ref{eqn:fisher:logn:typeI}) and (\ref{eqn:priors:logn}) into (\ref{eqn:mml87:codelength}) yields. As with the case of complete data, the MML87 estimates of $\mu$ and $\sigma$ must be obtained by numerical optimisation. We repeated the same model selection experiment as performed with complete data but this time varied the proportion of censoring from $10\%$ to $75\%$, similar to~\cite{KimYum08}. The performance of each method was measured in terms of probability of correct selection and the results are shown in Table~\ref{tab:results:mdl:typeI}. 

As shown in~\cite{KimYum08}, BIC performs better than SI, with the latter always preferring the Weibull distribution for large amounts of data. In terms of model selection accuracy, it is clear that the proposed MML87 method is superior to both BIC and SI, especially with small sample sizes or large amounts of censoring. Lastly, we note that MML codelengths derived in this paper can also be used in more complex applications such as mixture models and decision trees; for example, we may use the Weibull distribution to model data in the terminal nodes of a tree or the attributes of a class in a finite mixture model.  

\begin{table*}[btp]
\scriptsize
\begin{center}
\begin{tabular}{ccccccccccccccccc} 
\toprule
$n$ & $p$ & \multicolumn{3}{c}{${\bf y} \sim \text{Weibull}$} & & \multicolumn{3}{c}{${\bf y} \sim \text{Lognormal}$} & & \multicolumn{3}{c}{Average accuracy} \\
    &     & MML87 & BIC & SI & ~ & MML87 & BIC & SI & ~ & MML87 & BIC & SI \\
\multirow{4}{*}{25}  & 0.10 & {\bf  0.723} &  0.685 &  0.645 & &  0.853 &  0.828 & {\bf  0.854} & & {\bf  0.788} &  0.756 &  0.749\\ 
 & 0.30 & {\bf  0.613} &  0.563 &  0.494 & &  0.842 &  0.808 & {\bf  0.856} & & {\bf  0.728} &  0.686 &  0.675\\ 
 & 0.50 & {\bf  0.536} &  0.456 &  0.357 & &  0.833 &  0.795 & {\bf  0.866} & & {\bf  0.685} &  0.625 &  0.612\\ 
 & 0.75 & {\bf  0.603} &  0.289 &  0.146 & &  0.768 &  0.818 & {\bf  0.928} & & {\bf  0.685} &  0.554 &  0.537\\ 
\vspace{-2mm} \\ 
\cmidrule{2-13}
\vspace{-2mm} \\ 
\multirow{4}{*}{50}  & 0.10 & {\bf  0.830} &  0.810 &  0.790 & & {\bf  0.911} &  0.897 &  0.910 & & {\bf  0.870} &  0.853 &  0.850\\ 
 & 0.30 & {\bf  0.727} &  0.696 &  0.655 & &  0.875 &  0.852 & {\bf  0.880} & & {\bf  0.801} &  0.774 &  0.768\\ 
 & 0.50 & {\bf  0.629} &  0.582 &  0.532 & & {\bf  0.844} &  0.810 &  0.841 & & {\bf  0.736} &  0.696 &  0.686\\ 
 & 0.75 & {\bf  0.589} &  0.409 &  0.419 & & {\bf  0.822} &  0.785 &  0.744 & & {\bf  0.706} &  0.597 &  0.582\\ 
\vspace{-2mm} \\ 
\cmidrule{2-13}
\vspace{-2mm} \\ 
\multirow{4}{*}{100}  & 0.10 & {\bf  0.925} &  0.917 &  0.909 & & {\bf  0.963} &  0.958 &  0.963 & & {\bf  0.944} &  0.937 &  0.936\\ 
 & 0.30 & {\bf  0.839} &  0.822 &  0.812 & & {\bf  0.919} &  0.907 &  0.914 & & {\bf  0.879} &  0.865 &  0.863\\ 
 & 0.50 &  0.736 &  0.709 & {\bf  0.864} & & {\bf  0.876} &  0.854 &  0.521 & & {\bf  0.806} &  0.781 &  0.692\\ 
 & 0.75 &  0.625 &  0.525 & {\bf  0.954} & & {\bf  0.837} &  0.787 &  0.061 & & {\bf  0.731} &  0.656 &  0.508\\ 
\vspace{-2mm} \\ 
\cmidrule{2-13}
\vspace{-2mm} \\ 
\multirow{4}{*}{200}  & 0.10 & {\bf  0.983} &  0.981 &  0.979 & & {\bf  0.992} &  0.991 &  0.992 & & {\bf  0.987} &  0.986 &  0.986\\ 
 & 0.30 &  0.935 &  0.929 & {\bf  0.964} & & {\bf  0.968} &  0.964 &  0.896 & & {\bf  0.951} &  0.946 &  0.930\\ 
 & 0.50 &  0.847 &  0.834 & {\bf  0.998} & & {\bf  0.922} &  0.910 &  0.024 & & {\bf  0.884} &  0.872 &  0.511\\ 
 & 0.75 &  0.695 &  0.642 & {\bf  1.000} & & {\bf  0.854} &  0.815 &  0.000 & & {\bf  0.775} &  0.729 &  0.500\\
\bottomrule
\vspace{+1mm}
\end{tabular}
\caption{Probability of correctly selecting the data generating model for MML87, BIC and SI computed over $10^5$ simulation runs with type I censored data. The probability of censoring ($p$) is given in the second column.\label{tab:results:mdl:typeI}}
\end{center}
\end{table*}

\subsection{Type II censoring}
Our methodology can also be applied to data with type II censoring where the duration of the experiment is a random variable. In type II censoring, the experiment begins with $n$ items under observation and is stopped after the first $m$ failures are observed; clearly, with $m=n$ we have the case of complete data. The determinant of the expected Fisher information matrix for the Weibull distribution under Type II censoring is~\cite{WatkinsJohn06}
\begin{align}
\label{eqn:fisher:type2}
    |J(k,\lambda)| = \frac{m^2 \left(\pi ^2 -6 \phi _1^2+6 \phi _2\right)}{6 \lambda^2} ,
\end{align}
where
\begin{align}
    \phi_j = \frac{1}{m} \sum_{i=1}^m (-1)^{m-i} {n \choose i-1} {n-i-1 \choose m-i} \left(\log(n + 1 - i) \right)^j .
\end{align}
Note that $\phi_j = 0$ for complete data (ie, $m = n$) reducing (\ref{eqn:fisher:type2}) to the usual Fisher information for complete data (\ref{eqn:fisher:complete}). As with complete data and data with type I censoring, we  assume that the shape and scale parameters follow the half-Cauchy distribution a priori. To obtain MML estimates of the shape and scale parameters, we must again use numerical optimisation. Empirical experiments (not shown) demonstrate that the MML estimates are less biased and have better mean squared error compared to the maximum likelihood estimates in this setting.
%
%
\appendix

\section{Kullback--Leibler divergence between two Weibull distributions under type I censoring}
The Kullback--Leibler (KL) divergence~\cite{KullbackLeibler51} between the data generating model $p_0(y)$ and approximating model $p_1(y)$ is 
\begin{equation}
{\rm KL}( p_0 ||  p_1 ) = \int_{\mathcal{Y}} p_0(y) \log \frac{p_0(y)}{p_1(y)} dy ,
\end{equation}
where $\mathcal{Y}$ denotes the complete data space. Under the type I censoring setup described in Section~\ref{sec:censored} with a fixed censoring time $c > 0$, the KL divergence between two Weibull models Weibull($k_0, \lambda_0$) and Weibull($k_1, \lambda_1$) is
\begin{equation}
\label{eqn:weibull:kl}
{\rm KL}( k_0, \lambda_0 ||  k_1, \lambda_1) = \exp(-\left(c/\lambda_0\right)^{k_0}) A_1 + \left(\frac{\lambda_0 }{\lambda _1}\right)^{k_1} A_2 + \left(1-\frac{k_1}{k_0}\right) A_3 +\log \left(\frac{k_0}{k_1} \left(\frac{\lambda _1}{\lambda_0 }\right)^{k_1}\right)-1 ,
\end{equation}
where
\begin{align*}
A_1 &= \log \left(\frac{k_1 }{k_0}c^{k_1-k_0} \lambda_0^{k_0} \lambda_1^{-k_1}\right)+\left(\frac{c}{\lambda _1}\right){}^{k_1}+1  ,\\
A_2 &= \Gamma \left(\frac{k_1}{k_0}+1\right)-\Gamma \left(\frac{k_1}{k_0}+1,\left(\frac{c}{\lambda }\right)^k\right)  , \\
A_3 &= \text{Ei}\left(-\left(\frac{c}{\lambda_0 }\right)^{k_0}\right)-\gamma ,
\end{align*}
and $\text{Ei}(\cdot)$ is the exponential integral function
\begin{equation}
	\text{Ei}(z) = -\int_{-z}^\infty \frac{\exp(-t)}{t} \, dt .
\end{equation}
If the shape parameter of both densities is the same ($k=k_0=k_1$), the KL divergence simplifies to
\begin{equation*}
{\rm KL}( k, \lambda_0 ||  k, \lambda_1) = \left(1-\exp\left(-\left(\frac{c}{\lambda_0}\right)^k\right)\right) \left( \left(\frac{\lambda_0 }{\lambda _1}\right)^k+k \log \left(\frac{\lambda _1}{\lambda_0 }\right)-1\right)
\end{equation*}
The KL divergence between two Weibull densities with complete data can be obtained from (\ref{eqn:weibull:kl}) by noting that
\begin{align*}
	\lim_{c \to \infty} \exp(-\left(c/\lambda_0\right)^{k_0}) A_1 &= 0, \\
	\lim_{c \to \infty} \left( \frac{\lambda_0 }{\lambda_1} \right)^{k_1} A_2 &= \left(\frac{\lambda_0 }{\lambda _1}\right)^{k_1} \Gamma \left(\frac{k_1}{k_0}+1\right) , \\
	\lim_{c\to \infty} \left(1-\frac{k_1}{k_0}\right) A_3 &= \gamma  \left(\frac{k_1}{k_0}-1\right) 
\end{align*}
which, when substituted back into (\ref{eqn:weibull:kl}), simplifies to
\begin{equation*}
{\rm KL}( k_0, \lambda_0 ||  k_1, \lambda_1) = \left(\frac{\lambda_0 }{\lambda _1}\right)^{k_1}  \left(\frac{k_1}{k_0}-1\right) \gamma  + \log \left(\frac{k_0}{k_1} \left(\frac{\lambda _1}{\lambda_0 }\right)^{k_1}\right)-1   .
\end{equation*}
As an aside, by setting $k_0 = k_1 = 1$ in (\ref{eqn:weibull:kl}), we obtain the KL divergence between two exponential densities 
\begin{equation*}
p_T(t | \lambda_0) = \frac{\exp(-t/\lambda_0)}{\lambda_0}, \quad p_T(t | \lambda_1) = \frac{\exp(-t/\lambda_1)}{\lambda_1} ,
\end{equation*}
under type I censoring
\begin{equation}
{\rm KL}(  \lambda_0 ||   \lambda_1) = \left(1-\exp\left(-\frac{c}{\lambda_0 }\right)\right) \left(\frac{\lambda_0 }{\lambda _1}+\log \left(\frac{\lambda _1}{\lambda_0 }\right)-1\right) ,
\end{equation}
and for complete data
\begin{equation*}
{\rm KL}(  \lambda_0 ||   \lambda_1) = \frac{\lambda_0 }{\lambda _1}+\log \left(\frac{\lambda _1}{\lambda_0 }\right)-1.
\end{equation*}

\bibliographystyle{unsrtnat}
\bibliography{bibliography}  






\end{document}